\documentclass[twocolumn,preprint]{aastex63}
\usepackage{appendix}
\usepackage{amsmath,amstext}
\usepackage[T1]{fontenc}
\usepackage{apjfonts} 
\usepackage{natbib}
\usepackage{microtype}
\usepackage{longtable}
\usepackage{verbatim}
\usepackage{float}
\usepackage{booktabs}
\begin{document}
\title{Early Science with the Large Millimeter Telescope: Constraining the Gas Fraction of a Compact Quiescent Galaxy at $z=$1.883}
\author[0000-0001-9719-7177]{Joyce N Caliendo}
\affil{Physics Department, University of Connecticut,
    Storrs, CT 06269}
\author[0000-0001-7160-3632]{Katherine E Whitaker}
\affil{Department of Astronomy, University of Massachusetts, Amherst, MA 01003}
\affil{Cosmic Dawn Center, Copenhagen, Denmark}
\author[0000-0002-3240-7660]{Mohammad Akhshik}
\affil{Physics Department, University of Connecticut,
    Storrs, CT 06269}
\author[0000-0003-2222-4323]{Grant Wilson}
\affil{Department of Astronomy, University of Massachusetts, Amherst, MA 01003}

\author[0000-0003-2919-7495]{Christina C. Williams}
\affil{Steward Observatory, University of Arizona, Tucson, AZ 85721}
\affil{NSF Fellow}
\author[0000-0003-3256-5615]{Justin~S.~Spilker}
\affiliation{NHFP Hubble Fellow}
\affiliation{Department of Astronomy, University of Texas at Austin, 2515 Speedway, Stop C1400, Austin, TX 78712]}
\author[0000-0003-3266-2001]{
Guillaume Mahler}
\affiliation{Department of Astronomy, University of Michigan, 1085 S. University Ave, Ann Arbor, MI 48109, USA}
\author[0000-0001-8592-2706]{Alexandra Pope}
\affil{Department of Astronomy, University of Massachusetts, Amherst, MA 01003}
\author[0000-0002-7559-0864]{Keren Sharon}
\affiliation{Department of Astronomy, University of Michigan, 1085 S. University Ave, Ann Arbor, MI 48109, USA}
\author[0000-0002-7395-1988]{Emmaly Aguilar}
\affil{Corporaci\'{o}n Mexicana de Investigaci\'{o}n en Materiales S.A. de C.V., Ciencia y Tecnolog\'{i}a No 790 Col. Saltillo 400, C.P. 25290 Saltillo, Coahuila. M\'{e}xico}
\author[0000-0001-5063-8254]{Rachel Bezanson}
\affil{Department of Physics and Astronomy and PITT PACC, University of Pittsburgh, Pittsburgh, PA, 15260, USA}
\author[0000-0003-0407-8115]{Miguel Chavez Dagostino
}
\affil{Instituto Nacional de Astrof\'{i}sica \'{O}ptica y Electr\'{o}nica, Luis Enrique Erro 1, CP 72840, Tonantzintla, Puebla, M\'{e}xico
}
\author[0000-0001-9395-1670]{Arturo I. G\'{o}mez-Ruiz}
\affil{Instituto Nacional de Astrof\'{i}sica \'{O}ptica y Electr\'{o}nica, Luis Enrique Erro 1, CP 72840, Tonantzintla, Puebla, M\'{e}xico
}
\affil{ Consejo Nacional de Ciencia y Tecnolog\'{i}a, Av. Insurgentes Sur 1582, Col. Cr\'{e}dito Constructor, Alcald\'{i}a Benito Ju\'{a}rez, C.P. 03940, Ciudad de M\'{e}xico}

\author[0000-0003-4229-381X]{Alfredo Monta\~na}

\affil{Instituto Nacional de Astrof\'{i}sica \'{O}ptica y Electr\'{o}nica, Luis Enrique Erro 1, CP 72840, Tonantzintla, Puebla, M\'{e}xico
}
\affil{ Consejo Nacional de Ciencia y Tecnolog\'{i}a, Av. Insurgentes Sur 1582, Col. Cr\'{e}dito Constructor, Alcald\'{i}a Benito Ju\'{a}rez, C.P. 03940, Ciudad de M\'{e}xico}
\author[0000-0003-3631-7176]{Sune Toft}
\affil{Cosmic Dawn Center, Copenhagen, Denmark}
\affil{Niels Bohr Institute, University of Copenhagen, Jagtvej 128, DK-2200 Copenhagen, Denmark}
\author[0000-0003-1392-0159]{Miguel Velazquez de la Rosa}
\affil{Instituto Nacional de Astrof\'{i}sica \'{O}ptica y Electr\'{o}nica, Luis Enrique Erro 1, CP 72840, Tonantzintla, Puebla, México
}
\author[0000-0001-7737-2687]{Milagros Zeballos}
\affil{Universidad de las Am\'{e}ricas Puebla, Ex Hacienda Sta. Catarina M\'{a}rtir S/N. San Andr\'{e}s Cholula, Puebla, M\'{e}xico
}
\begin{abstract} We present constraints on the dust continuum flux and inferred gas content of a gravitationally lensed massive quiescent galaxy at $z$=1.883$\pm$0.001 using AzTEC 1.1mm imaging with the Large Millimeter Telescope. MRG-S0851 appears to be a prototypical massive 
compact 
quiescent galaxy, but has evidence that it experienced a centrally concentrated rejuvenation event in the last 100 Myr (see \citealt{mob}). This galaxy is undetected in the AzTEC image but we calculate an upper limit on the millimeter flux and use this to estimate the H$_2$ mass limit via an empirically calibrated relation that assumes a constant molecular gas-to-dust ratio of 150.
  We constrain the 3$\sigma$ upper limit of the H$_2$ fraction from the dust continuum in MRG-S0851 to be ${M_{\mathrm{H_2}}/M_{\mathrm{\star}}}$ $\leq$ 6.8\%.
  MRG-S0851 has a low gas fraction limit with a moderately low sSFR 
  owing to the recent rejuvenation episode, which together results in a relatively short depletion time of $<$0.6 Gyr if no further H$_2$ gas is accreted. Empirical and analytical models both predict that we should have detected molecular gas in MRG-S0851, especially given the rejuvenation episode; this suggests that cold gas and/or dust is rapidly depleted in at least some early quiescent galaxies.

\end{abstract}
\section{Introduction}
One of the biggest unanswered questions in galaxy evolution is what physical mechanism(s) cause galaxies to cease forming stars.  Many studies have been performed with the goal of better understanding what drives and halts star formation (e.g., \citealt{qqqq}). It has been established that there are two main types of galaxy populations spanning from the local to the distant Universe: less massive star-forming galaxies with blue colors owing to their young stars, and quiescent galaxies that are generally more massive with red colors due to older stellar populations  \citep{strat, whit2011}.  The most massive galaxies in the local Universe have predominantly old stellar populations, having formed most of their stars many Gyr ago \citep{mcderm15}.  By looking closer to their epoch of formation at $z>1$, massive recently quenched galaxies are a useful target population to search for observational signatures of quenching. 
\par The role of H$_2$ molecular gas depletion in quenching is not well understood \citep{jan2020}. Because quenching for most massive galaxies seems to occur on a rapid timescale, the number of galaxies experiencing this process at any time will be small \citep{schaw14, wild16}.  However, identifying these rare targets allows one to study the H$_2$ gas properties closest to the transition epoch, enabling investigations of poorly explored processes. There is a long list of physical mechanisms that could contribute to both halting and/or fueling star-formation in galaxies. In particular, molecular gas inflow and outflow may be a major factor in explaining their star formation histories. Gas heating and/or removal can impact quenching through processes like strangulation (e.g., \citealt{Peng15}), galactic merging events (e.g., \citealt{hop6}), halo quenching (e.g., \citealt{dek6}), 
and morphological quenching (e.g., \citealt{Martig9}).
\par Most star-forming galaxies follow the log(SFR)-log(M$_{\star})$ main sequence \citep{speagle14}, and as galaxies fall below this relation we observe a drop in the molecular gas fraction and star formation efficiency \citep{Tacc20}. Quantifying the gas reservoir and measuring its physical properties are important first steps towards developing an understanding of the role of cold gas and star-formation efficiency in driving quenching (e.g., \citealt{Saint17}). H$_2$ is difficult to detect directly, therefore two proxies are commonly used. One methodology includes using CO as a proxy; CO studies measure flux emitted by rotational transitions in the tracer molecules and relies on assumptions about the CO excitation and the CO to H$_{2}$ ratio. The second uses the dust continuum flux measurements assuming a dust temperature and H$_2$ to dust ratio. Here, we use dust to quantify molecular gas reservoirs and include comparisons to studies that utilize both methodologies. Quantifying the H$_2$ molecular gas reservoir is an important first step toward understanding what happens during the quenching process in massive high-$z$ galaxies.

\par In this letter, we place an upper limit on the H$_2$ reservoir of MRG-S0851, a strongly \color{black} gravitationally lensed massive galaxy at $z=1.88$ that is overall quiescent but experienced a burst of new star formation within the last 100 Myr. 
Our study contributes constraints on the H$_2$ gas content, as traced by 1.1mm dust continuum, associated with a rare example of a massive compact rejuvenating galaxy at $z\sim2$; currently, very few such measurements exist in the literature at \textit{z} \textgreater 1.5. We assume a standard $\Lambda$CDM cosmology with $\Omega_M$ = 0.3, $\Omega_\Lambda$ = 0.7 and H$_0$ = 70 km s$^{-1}$ Mpc$^{-1}$.

\begin{figure}[t]
    \includegraphics[width=\linewidth]{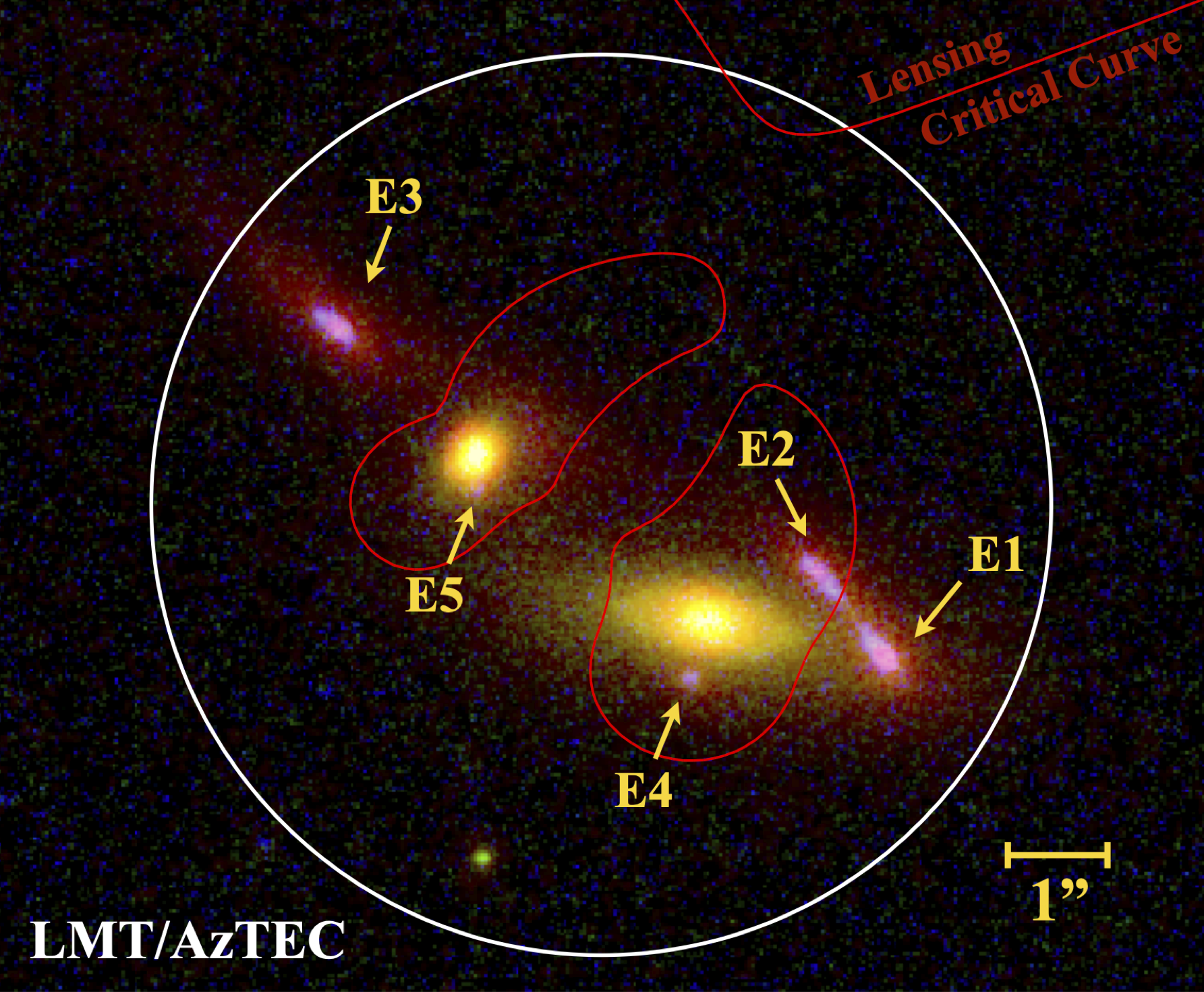}
    \caption{MRG-S0851 imaged by LMT/AzTEC. The composite color images are from HST/WFC3 H$_{\mathrm{F160W}}$
(red), I$_{\mathrm{F814W}}$ (green), and U$_{\mathrm{F390W}}$
    (blue).  There are five images within the $8\farcs5$ beamsize (white circle), but only E1, E2, and E3 were used in calculating the magnification factor, as E4 and E5 are negligible demagnified images. The red line is the critical curve, which marks regions in the image plane with theoretically infinite magnification. 
\label{fig:hstcolor}}
\end{figure}

\section{Data}
\label{sec:data}
\color{black}
Galaxy MRG-S0851 is a red galaxy at $z$=1.88 quintuply imaged via strong gravitational lensing \color{black} by the cluster SDSS~J0851$+$3331 identified in the Sloan Giant Arc Survey (Figure~\ref{fig:hstcolor}; see also \citealt{sharon20}). As a target in the REsolving QUIEscent Magnified (REQUIEM) galaxy survey, \citet{mo20} present the analysis of the stellar population gradients of our main target with surprising evidence supporting a rejuvenation phase in the last 100 Myr \citep{mob}.  MRG-S0851 is otherwise a prototypical $z\sim2$ quiescent galaxy, with an intrinsic stellar mass of $\log{M_*}/M{_\odot}$ = 11.02 $\pm$ 0.04, a low SFR of 6$\pm$1M$_{\odot}$/yr, sSFR log(SFR/M$_{\star}$) = -10.32$_{-0.05}^{+0.07}$ yr$^{-1}$, a S\'{e}rsic index of $n=2.35\pm0.05$, \color{black} and a compact circularized effective radius of 1.7$^{+0.3}_{-0.1}$ kpc. The SFR estimate of MRG-S0851 from \citet{mo20} was made using joint spectro-photometric fit assuming a non-parametric star formation history (SFH). These estimates are generally $\sim$0.1-1 dex lower than UV+IR estimates \citep[][see also Section~\ref{sec:lit}]{leja19}. Fitting the grism spectrum along with photometry decreases the SFR from $\sim$14 M$_{\odot}$/yr of photometry-only to $\sim$6 M$_{\odot}$/yr from the spectro-photometric fit.

MRG-S0851 was observed for 5.5 hours with the 1.1 mm continuum camera AzTEC on the Large Millimeter Telescope (Early Science Phase with 32m of active surface configuration) between February 12th and 22nd in 2016. The LMT is a 50 m single dish millimeter-wave radio-telescope located on the summit of Sierra Negra at an altitude of 4600 m above the sea level in the Mexican state of Puebla. AzTEC is 144 pixel bolometric camera with high mapping speed and high sensitivity which proves to be essential to observing  dust obscured galaxies \citep{Wilson8}.
The 1.1mm bandpass probes the Rayleigh-Jeans tail of the dust distribution in galaxies up to $z<$3.4, where the dust continuum has been shown to be a proxy for the molecular gas mass in star-forming galaxies \citep{scov16}.

\par The 5.5 hour integration was taken in good to excellent weather conditions ($\tau_{225GHz} = 0.04 - 0.07$).  Observations of the cluster SDSS~J0851+3331, centered on our target MRG-0851, were taken with the AzTEC small-map observing mode, covering an area of 7.5 arcmin$^2$.  The instrument has a point spread function (PSF) with a full width at half-maximum (FWHM) beam size of $8\farcs5$.  The fives images of MRG-S0851 span roughly $7\farcs0$ on the sky, as shown in Figure~\ref{fig:hstcolor}, such that the images are unresolved within the $8\farcs5$ LMT/AzTEC beam (white circle). Two (demagnified) images (E4 and E5) are negligibly faint in the optical, with the remaining three brighter images still resulting in a null detection in AzTEC. Full details about the lens model can be found in \citet{sharon20} and Appendix A of \citet{mo20}. The calibration and analysis of the AzTEC data follow the procedure described in \citet{Wilson8}. Our map covers 7.5 arcmin$^2$, with a mean RMS of 0.24 mJy.  We find excellent agreement between the 1$\sigma$ limits at the locations of E1-3, with an average value consistent at the $\pm$0.001 mJy level. There are no sources detected above an SNR > 4 within a 105" radius and we determine the 3 $\sigma$ upper limit on the 1.1mm flux from the RMS and the standard deviation of the weight map. 

\section{Methodology}

\begin{figure*}[!t]
\includegraphics[width=\linewidth]{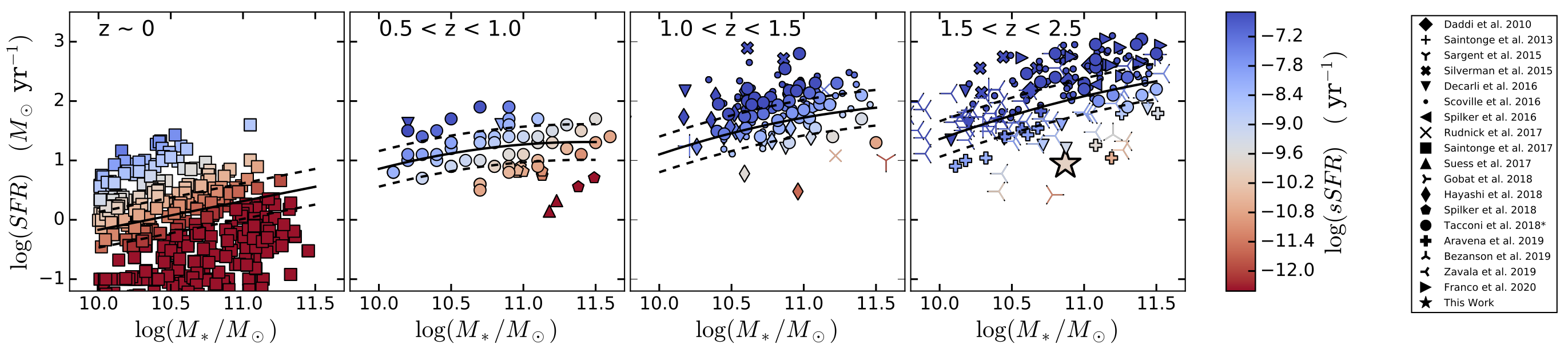}
\caption{This figure plots log(SFR) as a function of stellar mass.  Galaxies are color coded by their sSFR. The black lines represent the SFR-M${_*}$ main sequence from \citet{speagle14} for $z \sim $0 and from \citet{whit14} at $z > $0.5. The dotted lines are 0.3 dex above and below the main sequence. *See \citet{PHIBBS2} for full list of previous PHIBBS survey data included.
\label{fig:sfr}}
\end{figure*}

As noted in Section 2, our target MRG-S0851 went undetected in the deep 5.5 hour AzTEC/LMT integration. To calculate noise we found the RMS of the signal map and as a confirmation also compared to the standard deviation from the weight map.  We constrain the 3$\sigma$ upper limit of the dust continuum to be 0.72 mJy. At the redshift of MRG-S0851, the 1.1 mm imaging traces rest-frame 382$ \mu$m. Imaging at rest-frame wavelengths \textgreater 250 $\mu$m  to sample the Rayleigh Jeans tail where the dust continuum is a good proxy for the mass of the interstellar medium. At these wavelengths, the dust emission is optically thin so that the observed flux density is proportional to the mean temperature of dust, the mass of dust, and the dust opacity coefficient \citep{scov16}.  We assume a luminosity-weighted dust temperature of 25 K, which is the average temperature of galaxies at $z$ \textless 2.3 with little variation (see, e.g., \citealt{scov16}). In this methodology, the observed millimeter flux density is calibrated to estimate the molecular gas mass component of the ISM via observed galaxies in \citet{Scov14}.
Adopting our 3$\sigma$ upper limit, we use this dust continuum method to calculate a gas mass of $\log{M_{H_2}}$ = 10.69M$_{\odot}$, before correction for magnification for MRG-S0851, following
Equation 16 which is corrected in the erratum to \citet{scov16}. 
 
\par As the LMT beam size at the time of our observations was $8\farcs5$, all five images of our main target remain unresolved and therefore contribute to the null detection. 
Of these five images, we use three to calculate the magnification correction, as the other two are negligible demagnified images (see Figure~\ref{fig:hstcolor}). 
To determine the upper limit for a single image of MRG-S0581, we calculate the ratio of the total area in the HST image plane relative to the source plane to determine a magnification correction. Specifically, we define three regions in the image plane corresponding to E1, E2, and E3, and we map these regions back to the source plane using the lens model.  The ratio of the total projected area of these regions in the image plane relative to the source plane is calculated to be $\mu$=9.6, and is used as an estimate of the total magnification due to strong gravitational lensing.  As a confirmation, we also estimate the magnification using the \texttt{GALFIT} model \citep{Peng2} of images E1-3 (Appendix A; \citealt{mo20}), calculating a light-weighted average magnification of $\mu=9.46$; both estimates of magnification are therefore consistent. Flux contamination from the nearby cluster galaxies at rest-frame 800$\mu$m (observed 1.1mm at $z$=0.368) is negligible.   \color{black}

We calculate the upper limit on the intrinsic gas mass of MRG-S0851 by dividing the uncorrected measurement by the total magnification, finding <5.04$\times$10$^{9}$ M$_{\odot}$. This results in a 3$\sigma$ upper limit of the gas fraction for galaxy MRG-S0851 of $M_{\mathrm{H_2}}/M_{\mathrm{\star}}$<6.8\%.  Given the SFR corrected for systematic offsets (see Section~\ref{sec:lit}),
we therefore calculate a gas depletion timescale for MRG-S0851 to be M$_{\mathrm{H_2}}$/SFR < 0.6 Gyr.

\section{Comparison to Literature}
\label{sec:lit}
\color{black}
\par In order to compare to the LMT data presented herein, we conduct a literature search to collate a sample of galaxies with measured H$_2$ reservoirs and SFRs\footnote{Compilation available at \url{https://www.astrowhit.com/data-products}}. Below we summarize our composite reference sample, including the redshift and sample selection. The surveys use different SFR indicators owing to heterogeneous data sets. For all tracers, we homogenize measurements to a Chabrier IMF \citep{chab3} and an assumption of $\alpha_{\mathrm{CO}}$ = 4.4 where applicable. To trace H$_2$ gas, the surveys have either used CO emission lines or dust continuum, and target both individual galaxies and stacked samples. When using CO emission lines, all measurements are in lower transition levels (CO(2-1), CO(3-2) CO(1-0)). The adopted SFR indicators include: UV+IR SFRs (based on either 24$\mu$m empirical calibrations or low resolution far-infrared photometry), extinction corrected H$\alpha$ emission lines (or in one case [OII]), and SED fitting at ultraviolet to optical wavelengths, and in some cases extending to the MIR. Several of the surveys follow the \citet{wuyts} `SFR ladder', prioritizing first FIR photometry for high mass/star-forming galaxies, then empirically calibrated deeper 24$\mu$m as a proxy for SFRs for less massive galaxies, and finally parameterized SED modeling of the ultraviolet/optical light at the lowest sSFRs.

\par Most surveys of molecular gas in quiescent galaxies have been limited to the relatively nearby Universe.  Most notably, the xCOLD GASS survey did an extensive survey of galaxies at a redshift of $z\sim $0, including 532 galaxies selected to span a dynamic range in stellar mass and SFR \citep{Saint17}.  xCOLD GASS measurements of CO(1-0) are especially useful, as the sample serves as an unbiased reference to predict the behavior of the H$_2$ gas properties at higher redshift. 
\par At intermediate redshifts (0.5 $< z <$ 1.0) between 0.5 and 1.0, we include data from the following studies: PHIBSS2 survey by \citet{PHIBBS2}, \citet{Spilker18}, \citet{decarli2016}, and \citet{Suess17}.  Whereas the PHIBSS2 survey targets typical star-forming galaxies, \citet{Spilker18} and \citet{Suess17} both target intermediate redshift quiescent galaxies. 
\par For redshifts ranging from 1.0 \textless $z$ \textless 1.5, we include data from nine different studies. \citet{Sargent15, Hayashi18, decarli2016} and \citet{Daddi10}  are all studies using CO(2-1) to trace molecular hydrogen.  \citet{Sargent15} presented a measurement for one quiescent galaxy. \citet{Hayashi18} present 18 gas-rich galaxies at $z$ = 1.46, including one that is otherwise quiescent. \citet{Daddi10} include three star-forming galaxies. We include two starbursting galaxies from \citet{silverman15}. We include 37 star-forming galaxies from the PHIBSS2 survey \citep{PHIBBS2}. We also include galaxies from \citet{Saint13} and \citet{scov16}, who utilize data from the IRAM Plateau de Bure Interferometer and \textit{Herschel} PACS and SPIRE to study the dust and gas content. Finally, we include galaxies from the ASPECS survey \citep{ASPECS} and  GOODS-ALMA \citep{GOODS}.

\par For redshifts between 1.5 to 2.5, we include data from 14 different studies.  From \citet{Saint13, Daddi10, silverman15, decarli2016, scov16, GOODS, PHIBBS2} and \citet{ASPECS}, we include galaxies spanning all classifications.  Data collected from \citet{Spilker16} includes compact star-forming galaxies with low gas fractions. \citet{Rudnick17} includes cluster galaxies at $z$=1.62 with substantial gas reservoirs. \citet{Gobat18} present a stacking analysis of 977 quiescent galaxies at $z\approx1.8$ at far-infrared to sub-mm wavelengths by deblending \emph{Herschel} and SCUBA2 photometry to estimate the molecular gas content.  From \citet{Zav19} we include protocluster galaxies at $z$=2.1, some of which are gas poor and in a pre-quenching phase.  
Lastly, \citet{Bezanson} place a strong upper-limit on the H$_2$ gas mass of a quiescent galaxy at $z$  = 1.522.

\par Non-parametric SFHs are known to yield systematically lower sSFRs \citep{leja19}; to ensure an apples-to-apples literature comparison, we thus hereafter correct the stellar mass of MRG-S0851 by -0.15 dex and the SFR by +0.15 dex, yielding a +0.3 dex shift in sSFR following \citet{leja19}.  Even with this more conservative systematic correction, our conclusions remain unchanged.

\section{Results}
In Figure \ref{fig:sfr}, we show log(SFR) as a function of log(M$_{\mathrm{\star}}$) for four different redshift bins in the range $0.0<z<2.5$. 
Given the corrected specific SFR,  -9.9$_{-0.05}^{+0.07}$ yr$^{-1}$, MRG-S0851 is 1.0 dex below the average log(SFR)-log(M$_{\star}$) relation at $z=$1.88, and thereby classified as a quiescent galaxy.  As relatively few observations exist for galaxies below the main sequence at $z>1$, our new measurement contributes to poorly explored parameter space.

\begin{figure}
    \includegraphics[width=\linewidth]{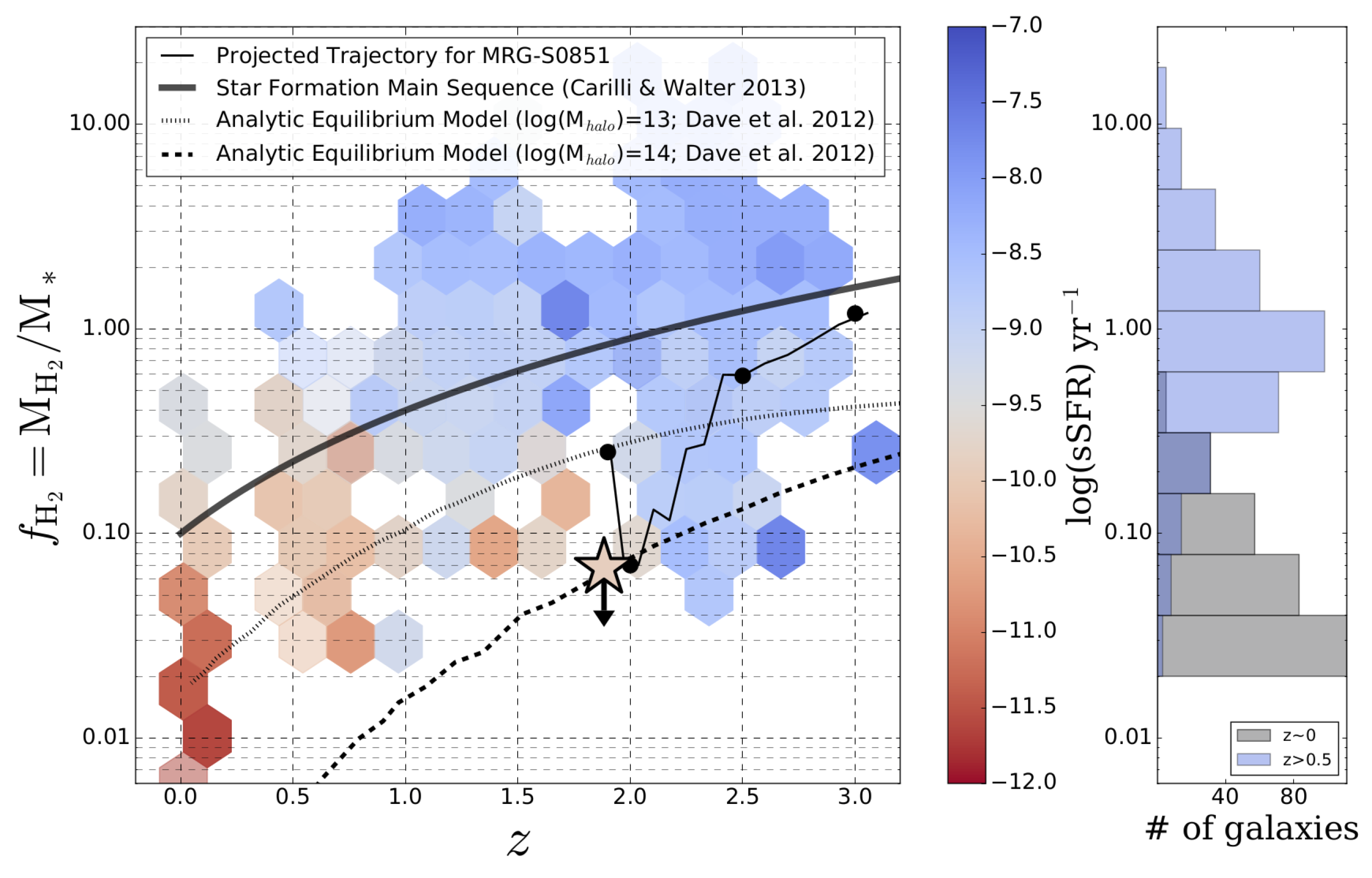}
\caption{This figure shows the gas fraction $M_{\mathrm{H_2}}/M_{\mathrm{\star}}$ as a function of redshift, color coded by the average log(sSFR) with a minimum of one galaxy per bin. Our target MRG-S0851 has the strongest limits placed to date for a quiescent galaxy at $z$>1.6. The thick black line is the gas fraction redshift main sequence adopted from \citet{Carilli13}. The thin black line represents the combination of the measured star formation history of MRG-S0851 with  empirical molecular gas scaling relations from \citet{PHIBBS2}. Comparisons to analytical equilibrium ``bathtub models'' by \citet{dave12} (dotted/dashed lines) demonstrate that only the most massive halos reach similarly low gas fractions by $z\sim2$.
\label{fig:fgas}}
\end{figure}

\begin{figure*}[t]
\centering
\includegraphics[width=0.85\linewidth]{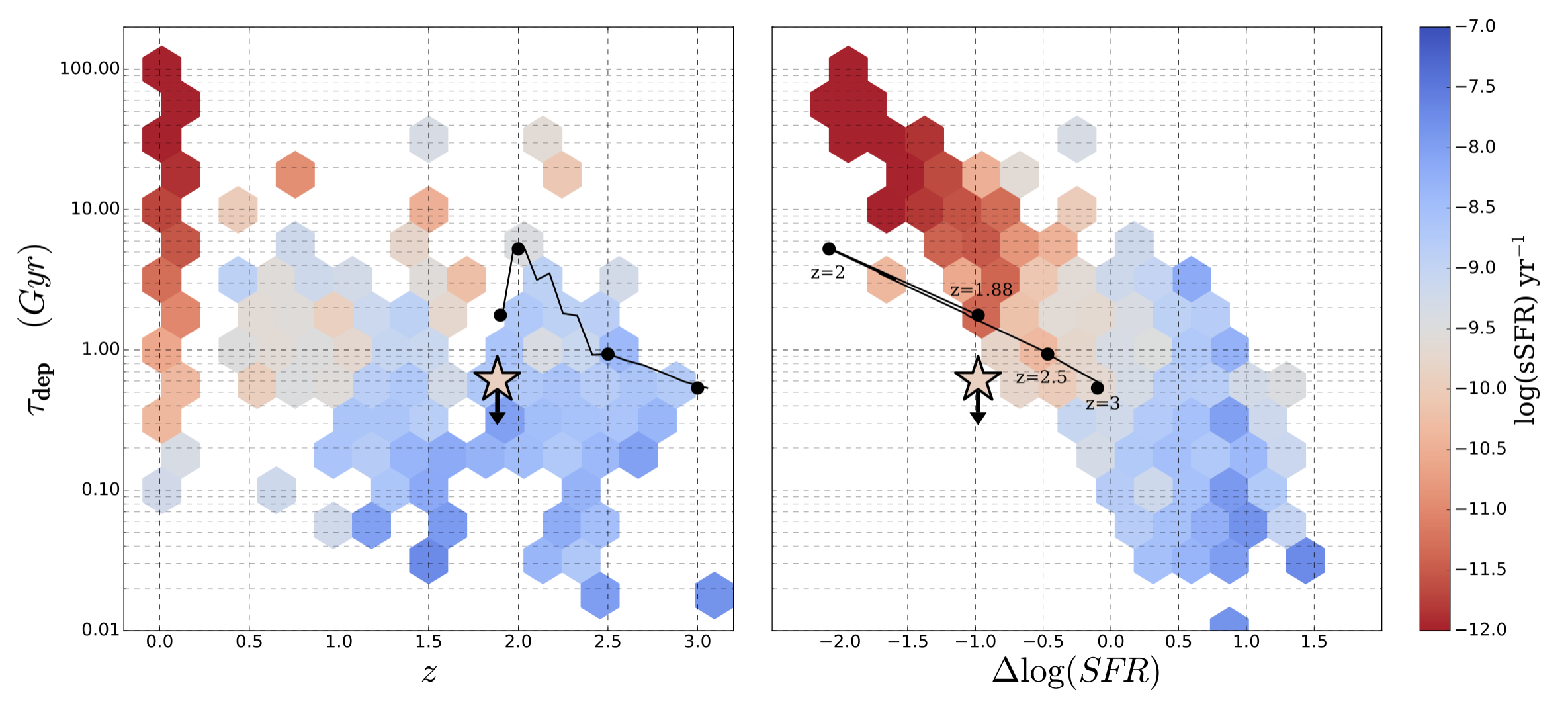}
\caption{Depletion timescale (M$_{\mathrm{H_2}}$/SFR) as a function of redshift (left) and distance from the average log(SFR)-log(M$_{\star}$) relation ($\Delta$log(SFR); right), color coded by sSFR. Evolutionary tracks for MRG-S0851 (thin black line) are inferred from the combination of the measured star formation history and empirical molecular gas scaling relations \citep{PHIBBS2} The depletion timescale for MRG-S0851 is an upper limit of 0.6 Gyr. The gas depletion timescale of this galaxy is relatively short, similar to star-forming galaxies at the same epoch (left), and a factor of 3 shorter than predicted by empirical relations.\color{black}}
\label{fig:depl}
\end{figure*}

\par  Figure~\ref{fig:fgas} shows the redshift evolution of the molecular gas fraction, defined as $f_{H_2}$ = ${M_{\mathrm{H_2}}/M_{\mathrm{\star}}}$. While the average $f_{H_2}$ in galaxies increases with redshift, MRG-S0851 has a relatively low gas fraction of $<$6.8\% compared to co-eval galaxies of similar stellar mass. For context, we compare to the predictions of analytical ``bathtub" models that balance gas inflow, outflow, and star formation in galaxies \citep[][see Section~\ref{sec:discussion} for more discussion]{dave12}. We additionally use the detailed star formation history for MRG-S0851 measured in \citet{mob} to explore its preceding evolution in gas fraction. The thin black line shows the gas fraction inferred using the empirical scaling relation of \citet{PHIBBS2} as a function of the projected past stellar mass and SFR of MRG-S0851, including systematic corrections as described in Section~\ref{sec:lit}. The late-time increase in gas fraction reflects the rejuvenation in the SFH, presenting a tension with our strong upper limit on $f_{H2}$ and suggesting relatively rapid gas consumption. 

\par The molecular gas depletion timescale, $M_{\mathrm{H2}}$/SFR, as a function of the redshift (left) and difference in SFR relative to the average log(SFR)-log(M$_{\star}$) relation ($\Delta$log(SFR); right) are shown in Figure~\ref{fig:depl}. Our value is 1.0 dex below the SFR-M$_{*}$ Main Sequence. 
While many quiescent galaxies have relatively low gas masses, they have similarly low SFRs which drive the longer depletion times.  
MRG-S0851 has a low gas fraction limit with an intermediate sSFR owing to the recent rejuvenation episode, which together results in a relatively short depletion time upper limit of 0.6 Gyr. Our measured short depletion time is in line with recent observations that suggest that scaling relations overestimate $f_{H2}$ and depletion time in the poorly-explored regime of low sSFR and high redshift \citep{Spilker18,ccw20}. 
\citet{mob} estimate that less than 1\% of the stellar mass (0.5$\pm$0.1\%)  formed  in the last 100 Myr as a result of this central rejuvenation.  Given the depletion timescale of $<$0.6 Gyr, if MRG-S0851 no longer accretes fresh H$_2$ gas, by $z\sim 1.6$ the new stars will have evolved and faded such that this target likely returns to resembling a compact, old quiescent galaxy.  
To agree with the general trend in the right panel of Figure~\ref{fig:depl}, where galaxies that lie below the average log(SFR)-log(M$_{\star}$) relation have significantly longer depletion times, the depletion timescale for MRG-S0851 would need to increase significantly with time.

\section{Discussion}
\label{sec:discussion}

\par Studying the molecular gas reservoirs in quiescent galaxies can provide insight into the formation and evolution of massive galaxies. While very few constraints exist for quiescent galaxies at $z>0$, there are observations of quiescent systems having surprisingly large H$_2$ reservoirs (e.g. \citealt{ Suess17, Gobat18}), however, this may not be the norm.  This study provides the strongest limits to date of a quiescent galaxy at $z>1.6$ using 1.1mm imaging with AzTEC. At $z=1.883$, we constrain the gas fraction of MRG-S0851 to $<$6.8\%.  
Results from similar galaxies range from $f_{gas}$ of $<$2-10\% \citep{ccw20, Bezanson, Sargent15, Zav19}, to 16\% \citep{Gobat18} to 40\% \citep{Rudnick17, Hayashi18}, indicating a diversity in molecular gas contents for high redshift quiescent galaxies.  

\par While the large scatter in observed molecular gas fractions may reflect the intrinsic variation within the quiescent population, it is also possible that the adopted methodologies are contributing to the apparent scatter. There are two differing methodologies for quantifying molecular gas and each have their respective uncertainties. In the case of dust tracing gas mass, \citet{Gobat18} perform the first statistical study of the H$_2$ gas properties of quiescent galaxies at a similar redshift to our target.  However, this study used far-IR to sub-mm data from the \emph{Herschel Space Telescope} and SCUBA2 whose large PSF complicates the analysis, with deblending issues lending considerable uncertainty to the analysis. While the dust continuum methodology can efficiently push to deeper sensitivity limits, uncertainties remain on the applicability of the calibrations presented in \citet{scov16} for low sSFR galaxies.  The calibrations in \citet{scov16} are based on data for galaxies in the local Universe, with studies showing that they hold on average for star-forming galaxies at $z\sim2$ \citep{k19}.  However, \citet{Li19} show that the dust-to-gas ratios at low sSFRs can vary dramatically (see also \citealt{privon18}). Perhaps the reason we find f$_{H_2}$ is low for MRG-S0851 is rather that the assigned dust-to-gas ratio is too high. While adopting the CO methodology would better constrain this ratio, uncertainties in the $\alpha_{\mathrm{CO}}$ conversion factor remain a source of much debate \citep{Hodge20, Tacc20}. 

\par MRG-S0851 has a rapid depletion time with an upper limit of 0.6 Gyr, about 10 times faster than similar quiescent galaxies at $z$=0. This implies that gas stabilization from collapse, a mechanism which has been invoked by \citet{Suess17} to explain high gas fractions, does not play a key role, at least for MRG-S0851. As discussed more extensively in \citet{ccw20}, the most massive halos (those that grow to $>10^{14}$ by $z=0$) could instead reach equivalently low gas fractions at $z\sim2$ by passing a critical mass threshold truncating gas accretion at early times. Abundance matching suggests MRG-S0851 is hosted in a dark matter halo of $\sim10^{12.6}$ M$_{\odot}$ at $z\sim2$ \citep{behroozi10}, which would naturally increase by a factor of 10 by $z\sim0$ (i.e., midway between the two analytical models in Figure~\ref{fig:fgas}).  Some tension of our low $f_{H2}$ measurement is therefore mitigated if MRG-S0851 resides in a more massive halo (e.g., dashed line in Figure~\ref{fig:fgas}), but only if the recently accreted gas is rapidly depleted.

\par It is interesting to note that the low gas fractions found by \citet{Spilker16} at $z\sim 2-2.5$ are for similarly compact galaxies (e.g., effective surface densities of 7.8x$10^9$-3.18x$10^{10}M_{\odot}$kpc$^{-2}$), but these targets are still actively forming new stars (see Figures~\ref{fig:fgas} and \ref{fig:depl}). The depletion times in \citet{Spilker16} are analogous to MRG-S0851 in the sense that they are both outliers among their star formation class. Perhaps their high stellar density increases the star-formation efficiency, as \citet{Bezanson} suggest.

\par The flat age gradient measured by Akhshik et al. (2020a) and the compactness of MRG-S0851 are consistent with an early-formation pathway. In the early-formation scenario, a galaxy forms the bulk of its stellar mass early in the Universe, consumes its gas and quenches \citep{will14, well15}. MRG-S0851 started forming stars between $z\sim$9 to $z\sim$3 and quenched by $z$=2. MRG-S0851's rejuvenation phase occurred between $z$=2 and $z$=1.88 (epoch of observation). Tentative evidence suggests that MRG-S0851 therefore accreted pristine gas at $z$=2, fueling a short-lived reignition. 

\par This letter presents only the first measurement of an individual quiescent galaxy at $z\sim2$ and we caution that this may not be representative of the quiescent population.  Continuing to study the molecular gas content of quiescent galaxies using a range of methodologies at high redshift is important both in understanding the basic gas property demographics of this population, as well as to test limitations in methodologies.  A larger sample size is also needed to determine how rapidly massive galaxies deplete and/or replenish their molecular gas and the mechanisms by which star formation halts at early times. The upcoming Legacy surveys with the TolTEC instrument on the 50m LMT will cover large extragalactic fields allowing for dust mass estimates for thousands of quiescent galaxies
out to $z\sim4$ with a spatial resolution of 5$\farcs$5. Specifically, the Ultra-Deep Survey will cover 0.8 sq. deg in the CANDELS fields to a detection limit of 0.1mJy (4$\sigma$) at 1.1mm.  We estimate that there are $\sim$2,000 quiescent galaxies above a stellar mass of  $\log{M_*}/M{_\odot}$ = 10 at $0<z<4$ in CANDELS, and we can constrain their gas fractions individually as low as 2\% for more massive galaxies or equivalently low through stacking analyses. Such a survey provides the required sensitivity and sample statistics to facilitate a revolution in our understanding of the role of gas fueling in truncating SF in massive galaxies. 

\begin{acknowledgements}
 We would like to thank the anonymous referee for sharing valuable comments that helped improve the letter significantly. K.W. wishes to acknowledge contributions made by Warren Sharpp to help shape this project.  We gratefully acknowledge  support by NASA under awards  No80NSSC19K1418, HST-GO-14622, and HST-GO-15663.
 This work is based on observations made with the NASA/ESA Hubble Space Telescope, obtained at the Space Telescope Science Institute, which is operated by the Association of Universities for Research in Astronomy, Inc., under NASA contract NAS 5-26555. Financial support for M.A. and K.W. is gratefully acknowledged. K.W. wishes  to  acknowledge funding from the Alfred P. Sloan Foundation. The  Cosmic Dawn Center is funded by the Danish National Research Foundation under grant No. 140. C.C.W. acknowledges support from the National Science Foundation Astronomy and Astrophysics Fellowship grant AST-1701546 and NIRCam Development Contract NAS50210 from NASA Goddard Space Flight Center to the University of Arizona.
 AM thanks the support from Consejo Nacional de Ciencia y Technolog\'ia (CONACYT) project A1-S-45680. 

\end{acknowledgements}

\end{document}